\documentclass[preprint,review,12pt,3p]{elsarticle}

\usepackage[numbers]{natbib}
\usepackage{graphicx}
\usepackage{dcolumn}
\usepackage{bm}
\usepackage[squaren, Gray, cdot]{SIunits}
\usepackage[utf8]{inputenc}
\usepackage[T1]{fontenc}
\usepackage{mathptmx}
\usepackage{xcolor,soul}
\usepackage[normalem]{ulem}
\usepackage{amsmath}
\usepackage{booktabs} 

\journal{arXiv}

\begin{document}
\begin{frontmatter}

\title{Magnetic properties of Nd$_6$Fe$_{13}$Cu single crystals}

\author[1]{Jianing Liu\corref{cor1}}
\ead{jianing.liu@tu-darmstadt.de}

\cortext[cor1]{Corresponding author}

\author[2]{Ruiwen Xie}%

\author[1]{Alex Aubert}%

\author[1]{Lukas Sch{\"a}fer}%

\author[2]{Hongbin Zhang}%

\author[1]{Oliver Gutfleisch}%

\author[1]{Konstantin Skokov}%

\address[1]{Functional Materials, TU Darmstadt, 64287 Darmstadt, Germany}
\address[2]{Theory Magnetic Materials, TU Darmstadt, 64287 Darmstadt, Germany}

\begin{abstract}
The understanding of coercivity mechanism in high performance Nd-Fe-B permanent magnets relies on the analysis of the magnetic properties of all phases present in the magnets.
By adding Cu in such compounds, a new Nd$_6$Fe$_{13}$Cu grain boundary phase is formed, however, the magnetic properties of this phase and its role in the magnetic decoupling of the matrix Nd$_2$Fe$_{14}$B grains are still insufficiently studied. In this work, we have grown  Nd$_6$Fe$_{13}$Cu single crystals by the reactive flux method and studied their magnetic properties in detail. It is observed that below the N{\'e}el temperature \mbox{($T_N=410$~K)}, the Nd$_6$Fe$_{13}$Cu is antiferromagnetic in zero magnetic field; whereas when a magnetic field is applied along the a-axis, a spin-flop transition occurs at approx. 6~T, indicating a strong competition between antiferromagnetic and ferromagnetic interactions in two Nd layers below and above the Cu layers. Our atomistic spin dynamics simulation confirms that an increase in temperature and/or magnetic field can significantly change the antiferromagnetic coupling between the two Nd layers below and above the Cu layers, which, in turn, is the reason for the observed spin-flop transition. These results suggest that the role of antiferromagnetic Nd$_6$Fe$_{13}$Cu grain boundary phase in the coercivity enhancement of Nd-Fe-B-Cu magnets is more complex than previously thought, mainly due to the competition between its antiferro- and ferromagnetic exchange interactions.
\end{abstract}


\end{frontmatter}

 \pagebreak 
Modern Nd-Fe-B-based permanent magnets are one of the most powerful hard magnetic materials that dominate today’s market ~\cite{gutfleisch2000controlling,skokov2018heavy,sugimoto2011current}. The microstructure of a high-performance Nd-FeB magnet should consist of micro- or nanometer-sized Nd$_2$Fe$_{14}$B hard magnetic grains, isolated (i.e.; magnetically decoupled) by a paramagnetic Nd-rich grain boundary phase ~\cite{sawatzki2018anisotropic,schafer2022novel,liu2015grain,opelt2021upscaling}, regardless of the synthesis techniques.  Nevertheless, it is known that besides such a ‘classical’ microstructure, Nd-Fe-B-based magnets can contain additional magnetic phases located either inside the 2:14:1 grains \cite{sepehri2019development} or as grain boundary phases ~\cite{schafer2021design, goll2021additive}, and the coercivity of such magnets remains at a high level.

It is known that a small amount of Cu or Ga improves the wettability of grain boundary rare-earth rich phase in the Nd-Fe-B-based permanent magnets, showing a pronounced effect on increasing the coercivity ~\cite{li2009role,herbst1991r,shimoda1988high}. At the same time, a new grain boundary phase Nd$_6$Fe$_{13}$Cu (or Nd$_6$Fe$_{13}$Ga) is present ~\cite{xu2019comparison,sasaki2016formation}. However, the exact magnetic state of this grain boundary phase (paramagnetic, ferromagnetic or antiferromagnetic) and its role in the formation of a high coercive state are still under discussion \cite{huang2022enormous,zhang2022coercivity,zhang2022effect}. This situation is mostly related to the lack of information in the literature on the intrinsic magnetic properties of the 6:13:1 phase.

The system RE$_6$Fe$_{13-x}$TM$_x$(RE = Nd, Pr, Sm; TM = Cu, Ag, Au, Si, Ge, Al, Sn, Ga) was investigated by many groups in the past~\cite{hu1992crystal,yan1994magnetic,kajitani1992microstructure,weitzer1994magnetism,knoch1993structure}. Still, there is a strong debate on the magnetic order of the 6-13-1 phase. Kajitani, de Groot and Goll reported the compound to be antiferromagnetic~\cite{goll2021additive,kajitani1992microstructure,de1998magnetic}; whereas Hu and Yan stated that it has to be ferrimagnetic \cite{hu1992crystal,yan1994magnetic}, and Weitzer reported it to be ferromagnetic~\cite{weitzer1994magnetism}.  Although the magnetic order is discussed for various compositions, the Nd$_6$Fe$_{13}$Cu (or Pr$_6$Fe$_{13}$Cu) phase is the most interesting for permanent magnet applications, nevertheless, the information about the magnetic properties of this compound is sometimes controversial. Knoch studied  Nd$_6$Fe$_{13}$Cu polycrystalline powder and reported it with a ferrimagnetic spin structure. Furthermore, it was reported that there was a spin reorientation transition upon cooling at around 4.2~K from ferri- to ferromagnetic order~\cite{knoch1993structure}. Hautot investigated the Nd$_6$Fe$_{13}$Cu compound and reported it as antiferromagnetic order with the N{\'e}el temperature ($T_N$) at 408~K~\cite{hautot1998mossbauer}, whereas for the same composition, Kajitani reported $T_N$ = 420~K ~\cite{kajitani1992microstructure}, Knoch claimed that $T_C$ = 463~K ~\cite{knoch1993structure} and Weitzer showed $T_C$ = 430~K ~\cite{weitzer1994magnetism}. Such discrepancies can be explained by the fact that the magnetic measurements in the literature were done on polycrystalline samples, containing a small amount of Nd$_2$Fe$_{17}$ phase with $T_C$ = 327~K \cite{diop2018origin}.

From theory, first-principle calculations show that the hypothetical binary Nd$_6$Fe$_{14}$ compound has a ferromagnetic ordering (FM), while the substitution of Ga for Fe atoms in Nd$_6$Fe$_{13-x}$Ga$_X$ makes the antiferromagnetic (AFM) state stable due to the anti-parallel coupling between the neighbouring Nd-Fe blocks separated by Ga atoms \cite{jonen1997metamagnetic}. This AFM-FM competition leads to a metamagnetic transition occurring at low temperatures in the Nd$_6$Fe$_{13-x}$TM$_X$ compounds with TM=Ga \cite{zhao1995antiferromagnetic}, Al \cite{jonen1997metamagnetic}, Au \cite{de1996erratum}, Sn \cite{rosenberg1994metamagnetism}, Pd \cite{ruzitschka2002investigation}, where a relatively high external magnetic field can change the magnetic ordering from the AFM to the FM states. The presence of a metamagnetic transition has also been reported for the Nd$_6$Fe$_{13}$Cu phase, but the critical field of such a metamagnetic transition varies from 1.5~T ~\cite{knoch1993structure} to 10~T~\cite{de1998magnetic}, and sometimes the abrupt change in magnetization was interpreted as a spin reorientation transition caused by temperature dependencies of anisotropy constants of Nd-ions \cite{iranmanesh2011magnetoelastic}. It is known that the critical field of metamagnetic transition strongly depends on the orientation of the crystals with respect to the external magnetic fields; however, none of these works have studied the magnetic properties in Nd$_6$Fe$_{13-x}$TM$_X$ single crystals, which is necessary for an unambiguous determination of the intrinsic magnetic properties.

In this study, to investigate precisely the intrinsic magnetic properties of Nd$_6$Fe$_{13}$Cu, we grow single crystals of 1x1x0.1 mm in size and measure the magnetic properties along different crystallographic directions in a wide range of external magnetic fields and temperatures. It is shown that Nd$_6$Fe$_{13}$Cu has an AFM structure with a N\'eel temperature of 410~K, which is further confirmed by our atomistic spin dynamics (ASD) simulation predicting an antiferromagnetic ground state with a N\'eel temperature of around 440 K. 
At temperatures below 150~K, an external magnetic field of 6-7 T applied along the c-axis induces a spin-flop transition, indicating a strong competition between the antiferro- and ferromagnetic exchange interactions between the Nd and Fe sublattices. Our experimental results are supported by the atomistic spin dynamics simulation showing that the increasing temperature and/or magnetic field provide extra driving forces to compete with the antiferromagnetic exchange couplings between the two Nd layers below and above the Cu layers, which, in turn, is the reason for the observed spin-flop transition. Moreover, the competing behaviour in the exchange couplings can be well interpreted according to the calculated exchange coupling parameters based on density functional theory (DFT).

The reactive flux technique was applied to grow Nd$_6$Fe$_{13}$Cu single crystals \cite{latturner2018clusters,kanatzidis2005metal}. The Nd$_6$Fe$_{13}$Cu phase crystallizes in the peritectic reaction Liq.+Nd2Fe17 $\longrightarrow$ Nd$_6$Fe$_{13}$Cu at $\sim$900~K, and turned out to be stable below this temperature. According to the ternary phase diagram, we designed the optimal composition Nd$_{73}$Fe$_9$Cu$_{18}$ and prepared the master bulk alloy by induction melting under argon atmosphere. The obtained ingot was broken into pieces, placed in a ZrO$_2$ crucible and sealed in quartz ampule under vacuum. The heat treatment consists of dwelling the sample at 600~\degreecelsius~for 1h, then a slow cooling (15~K/h) to 570~\degreecelsius~, then further cooling with 0.5~K/h to 500~\degreecelsius~with subsequent quenching in water. In order to check the stability of the Nd$_6$Fe$_{13}$Cu phase at temperatures used for the heat treatment of the Nd-Fe-B magnets, several single crystals were selected, wrapped in tantalum foil, sealed in quartz tube and then annealed by heating up from room temperature to 600~\degreecelsius~(300~K/h), dwelled at this temperature for 1~h and cooling with furnace to room temperature~\cite{schafer2021design, goll2021additive}.

To analyze the microstructure, phase analysis was performed using a Tescan VEGA 3 Scanning Electron Microscope (SEM) with a backscattered electron (BSE) detector. In addition, energy dispersive x-ray spectroscopy (EDX) was performed to quantify the local composition. The quality of the grown crystals was checked with a backscattering Laue diffractometer. Suitable crystals were oriented for measurements of the intrinsic magnetic properties. Temperature and field dependencies of magnetization were performed using a PPMS-VSM (Quantum Design PPMS-14) and Magnetic Property Measurement System (Quantum Design MPMS).

To understand the spin texture evolution with respect to temperature, as well as with respect to external magnetic fields, we performed atomistic spin dynamics (ASD) simulations using the Uppsala Atomistic Spin Dynamics (UppASD) software~\cite{eriksson2017atomistic}, implemented based on the Landau-Lifshitz-Gilbert (LLG) equation. 
As inputs for the ASD simulations, the exchange coupling parameters $J_{ij}$ were calculated using the post-processing code `jx'~\cite{terasawa2019efficient} provided by OpenMX~\cite{ozaki2003variationally} after an LDA + U self-consistent calculation~\cite{han2007magnetic}.  
In the self-consistent calculation, the energy cutoff and energy convergence criteria were set to 400 Ry and 1.0$\times$10$^{-8}$ Hartree, respectively. 6$\times$6$\times$6 and 8$\times$8$\times$8 k-meshes were adopted for self-consistent and $J_{ij}$ calculations, respectively.

\begin{figure}[h!]
\centering
\includegraphics[height=0.25\textwidth]{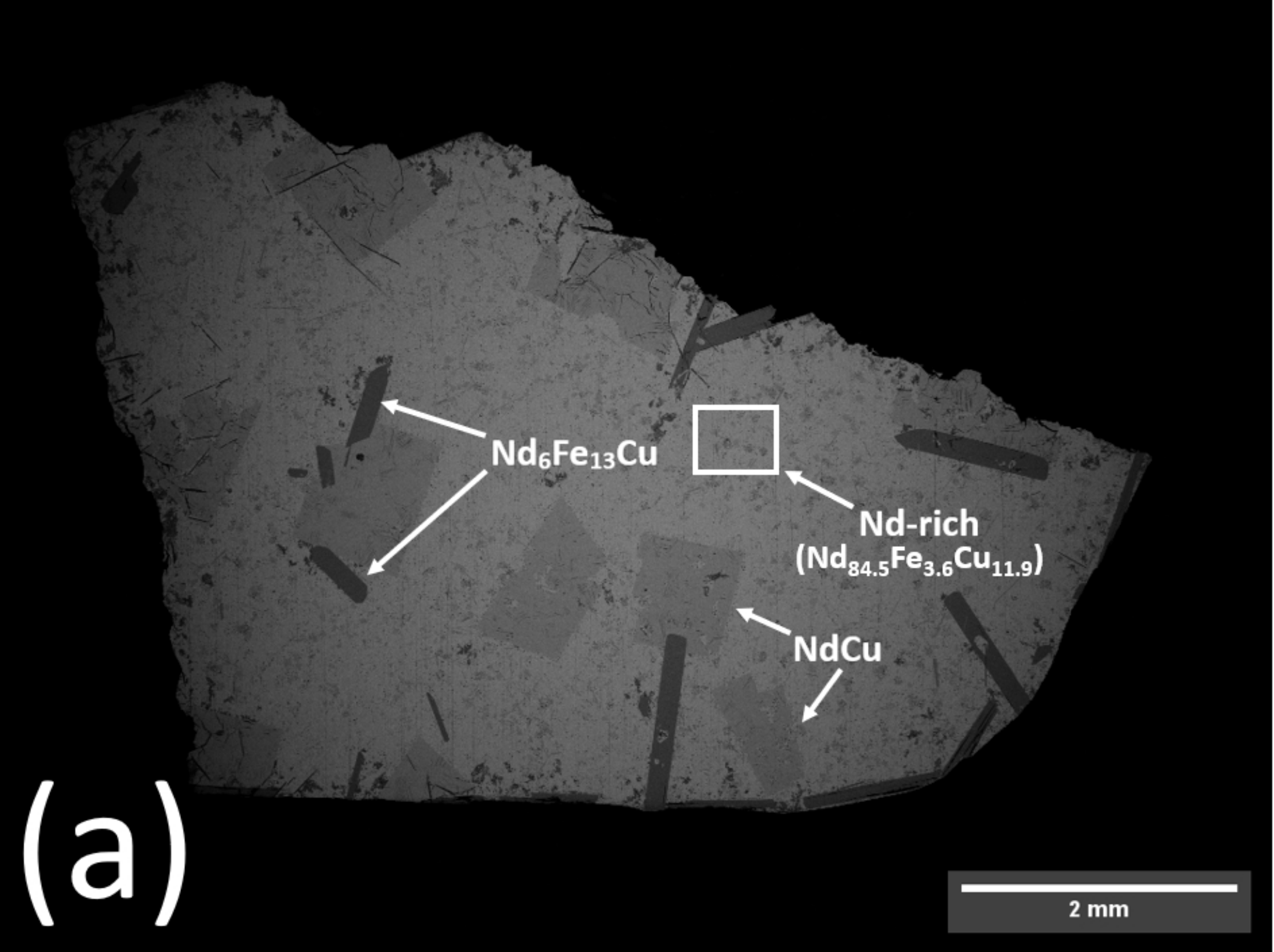}
\includegraphics[height=0.25\textwidth]{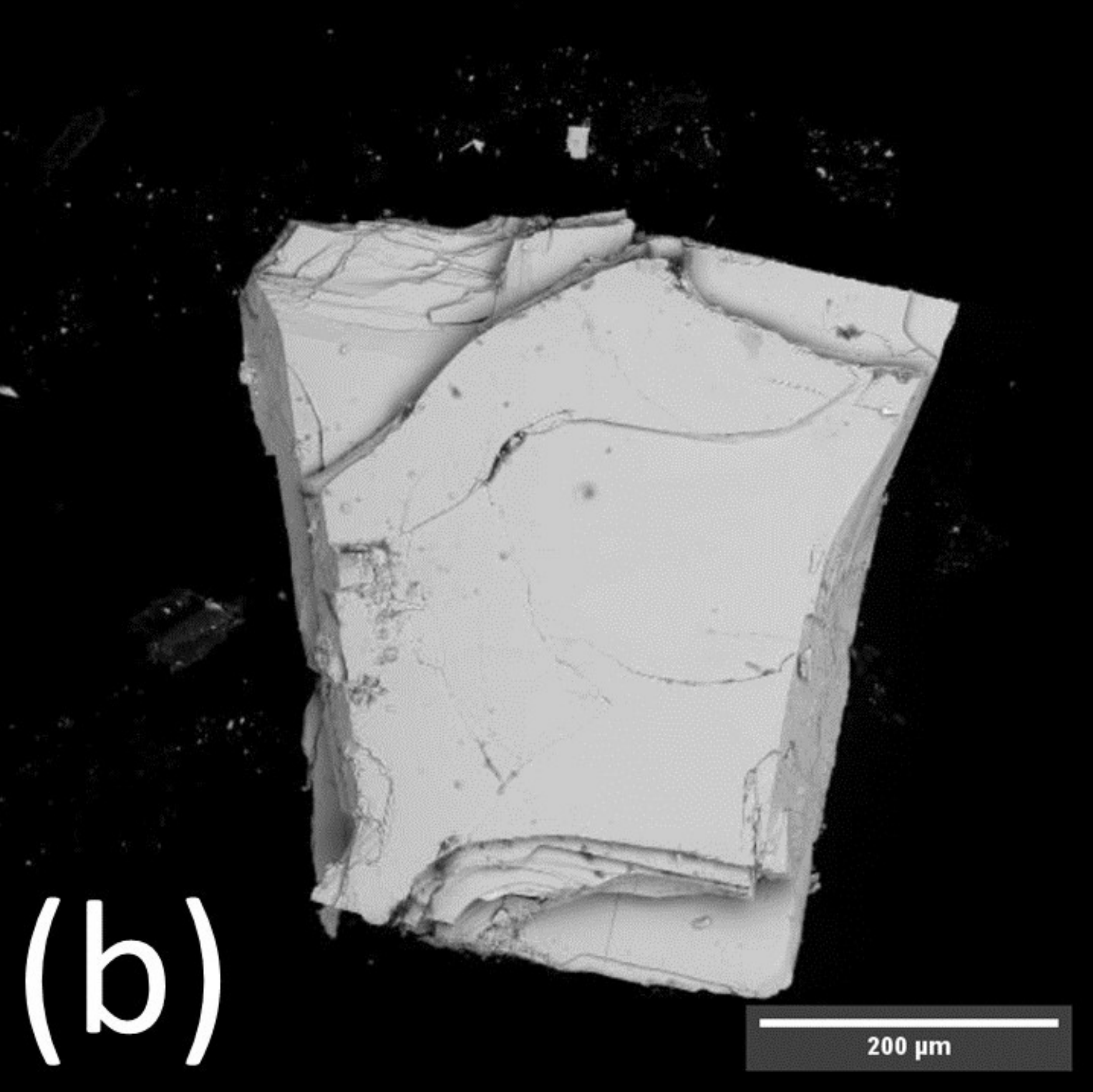}
\includegraphics[height=0.25\textwidth]{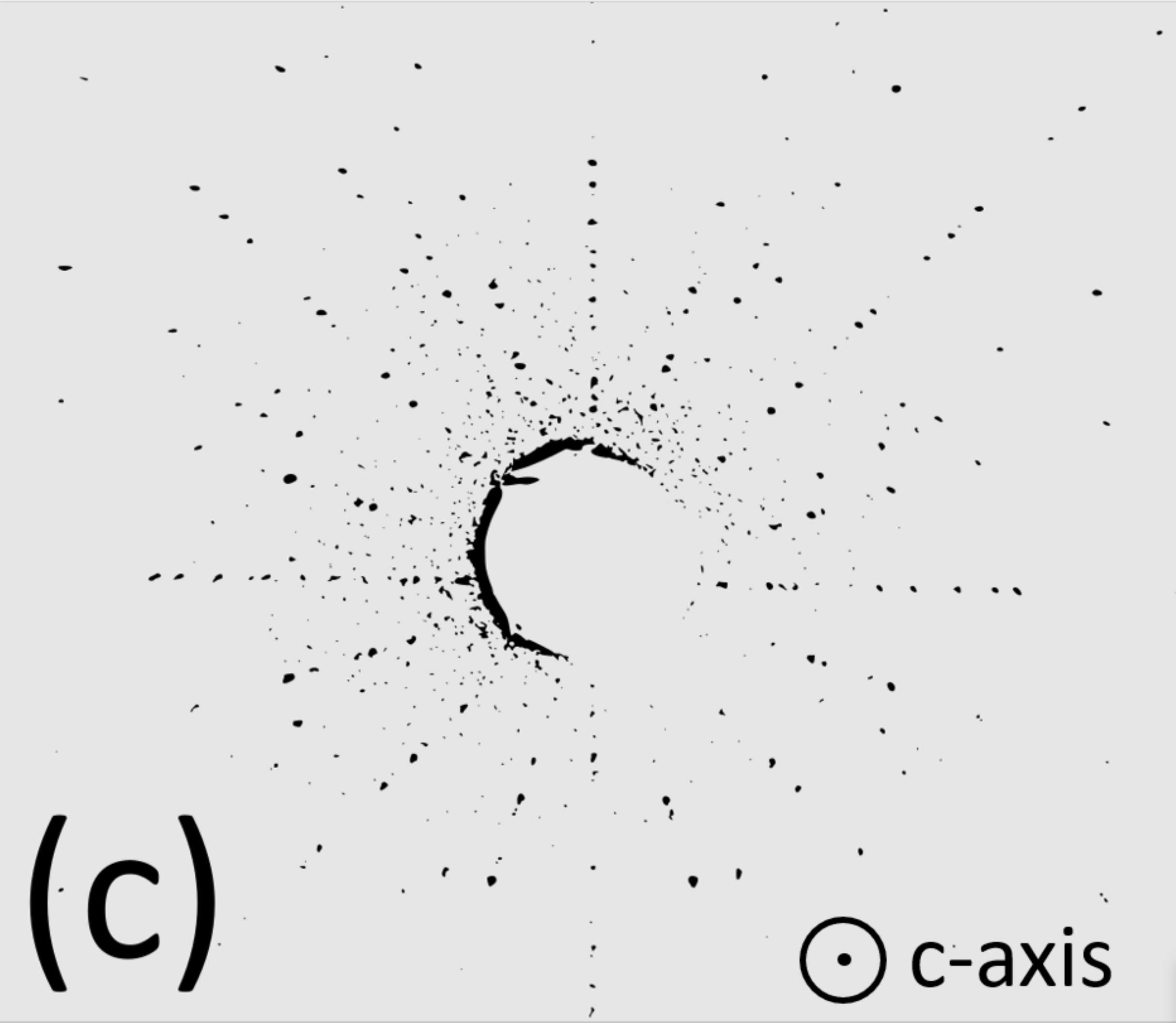}
\caption{(a) BSE image of the ingot after single crystal growing, (b) extracted single crystal of Nd$_6$Fe$_{13}$Cu phase, and (c) Laue X-Ray back-scattered diffraction of Nd6Fe13Cu single crystal. }
\label{SC}
\end{figure}

Fig.~\ref{SC}~(a) shows the BSE image of the Nd$_{73}$Fe$_9$Cu$_{18}$ alloy after the single crystal grow. The ingot contains three different phases: oblong grains of desired Nd$_6$Fe$_{13}$Cu (dark), which were extracted by short time etching in water solution of cytrick and acetic acid (2\%-4\%). Other phases which are shown in Fig.~\ref{SC}~(a) are square-like NdCu phase (gray) and the Nd-rich phase (light, as an example, the composition of the area marked in square is  Nd$_{84.5}$Fe$_{3.6}$Cu$_{11.9}$).  Fig.~\ref{SC}~(b) shows the BSE image of an extracted single crystal of Nd$_6$Fe$_{13}$Cu phase. It can be seen that the Nd$_6$Fe$_{13}$Cu crystal has a layered structure, but the mechanical stability of the crystal is good. The backscattered Laue diffraction was performed to ensure the quality of the crystal and to determine the orientation of the principal crystallographic direction. Fig.~\ref{SC}~(c) shows a diffraction pattern obtained from a flat surface of the crystal (out-of-plane), showing that this orientation is (001).

\begin{figure}[h!]
\centering
\includegraphics[width=1\textwidth]{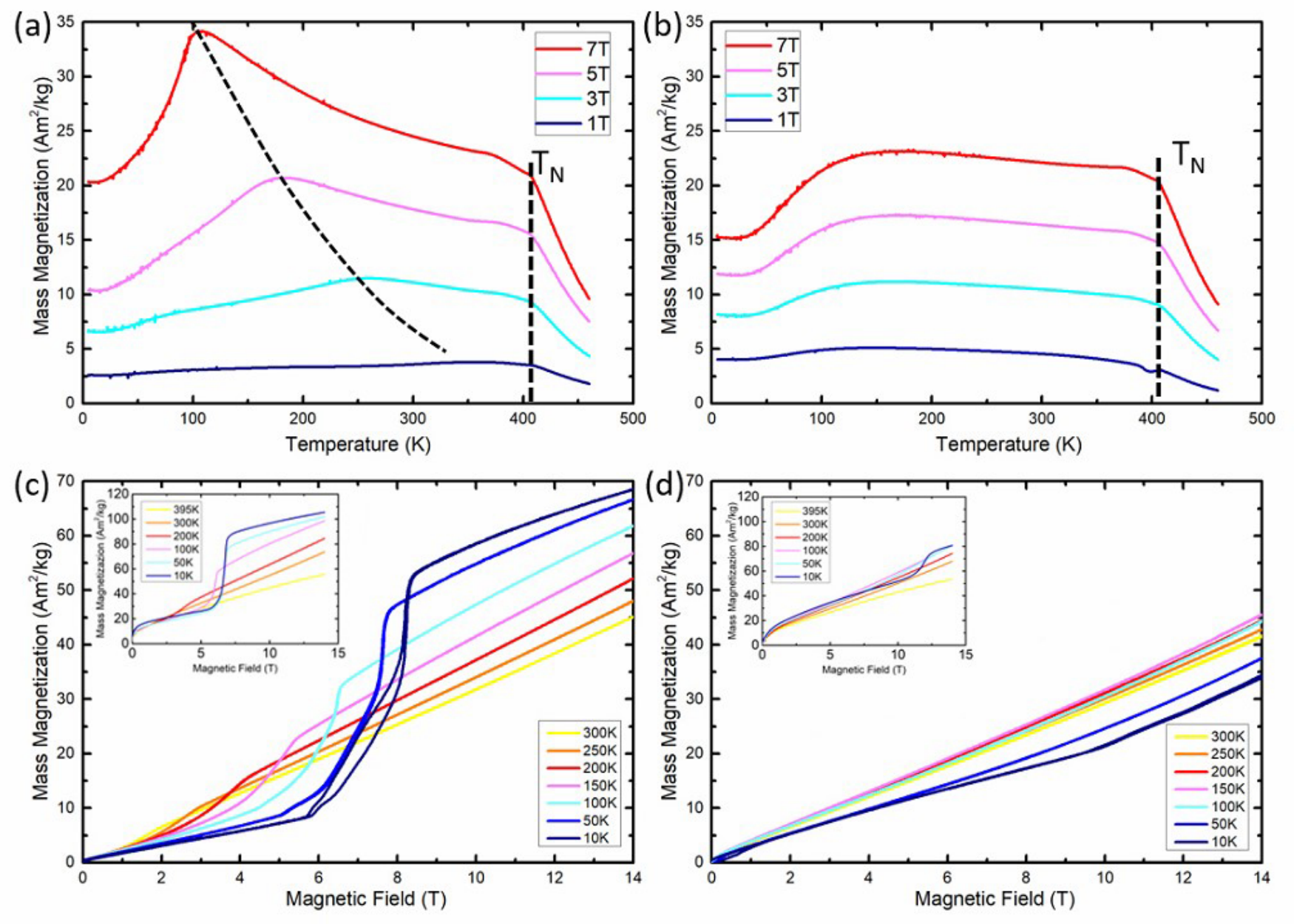}
\caption{(a) a-axis (in-plane) and (b) c-axis (out-of-plane) $M(T)$ curves measured under different external fields (1, 3, 5, and 7~T); and  (c) a-axis (in-plane) and (d) c-axis (out-of-plane) $M(H)$ curves. The inserts in (c) and (d) show $M(H)$ curves measured on additionally annealed single crystal.}
\label{MTMH5}
\end{figure}

The magnetic properties of Nd$_6$Fe$_{13}$Cu single crystal were measured in two different directions: in-plane (along the a-axis) and out-of-plane (along the c-axis). Fig.~\ref{MTMH5}(a) shows the temperature dependencies of magnetization measured along the $a$ direction in magnetic fields 1, 3, 5 and 7~T. A kink around 410~K corresponds to the N\'eel temperature $T_N$. Similar results were also shown in the work of O. Isnard, where instead of Cu the Si was used to stabilize 6:13:1 phase~\cite{isnard2002neutron}. In our case, in the temperature range below the $T_N$, the $M(T)$ curves of Nd$_6$Fe$_{13}$Cu single crystal demonstrate non-typical behavior for the ‘pure’ antiferromagnetic phase: instead of a continuous decrease in the magnetization with the temperature, the magnetic moment of the sample first rises under cooling, indicating that an inter-sublattice antiferromagnetic exchange interaction weakens with decreasing temperature. The $M(T)$ curves show a well-defined second kink at temperatures below 200~K, which gradually shifts towards low temperatures in a high magnetic field (a dotted line is added to the figure to show the evolution of this transition with the magnetic field).

Fig.~\ref{MTMH5}(b) shows the $M(T)$ curves measured along the $c$-axis of the crystal in magnetic fields of 1, 3, 5 and 7~T. It can be seen that at low temperatures the $M(T)$ curves measured in a-axis (in-plane) differ greatly from the c-axis (out-of-plane) $M(T)$ dependencies, showing that the magnetic properties of Nd$_6$Fe$_{13}$Cu are highly anisotropic below $T_N$. The low-temperature kink in the $M(T)$ curves measured along the $a$-axis is still visible, but it is not as pronounced as in the case of a magnetic field applied along the $c$-axis.

The field dependencies of magnetization $M(H)$ measured in a magnetic field up to 14 T at various temperatures are shown in Fig.~\ref{MTMH5}(c) (the magnetic field is along $c$-direction ) and in Fig.~\ref{MTMH5}(d) ($H$ is parallel to the crystallographic axis $a$). It can be seen that when the magnetic field is applied along the $c$-direction, at temperatures below 250~K a field-induced spin-flop transition occurs, and the jump in magnetization becomes more noticeable at temperatures below 150~K. At 10~K the spin-flop transition is accompanied by magnetic hysteresis, which indicates the first-order nature of the transition. In contrast, when the magnetic field is applied along the $a$-axis, there is no spin-flop transition in the magnetic field up to 14~T, and the magnetization gradually increases with the field, demonstrating typical magnetization behavior of an antiferromagnetic material. Thus, we can suppose that the antiferromagnetic exchange interactions depend strongly on the crystallographic orientation and that the magnetic anisotropy of Nd ions seems to play an important  role in the formation of the antiferromagnetic structure.

\begin{table}[h]
\centering
\caption{Summary of magnetic properties of Nd$_6$Fe$_{13}$Cu single crystals.}
\begin{tabular}{cccc}
\toprule  
&  M(14T) at 300K ($Am^2$/$Kg$)  &  M(14T) at 10K ($Am^2$/$Kg$)  & $T_N$ (K)\\
\midrule 
a-axis & 45.11 & 68.47 & 410\\
c-axis & 41.39 & 34.40 & 410\\
\bottomrule 
\end{tabular}
\label{table}
\end{table}

Obviously, the magnetic state of Nd$_6$Fe$_{13}$Cu phase in the field up to 6~T is always antiferromagnetic, and there is no experimental evidence for ferrimagnetic or canted magnetic structures in the low fields. At the same time, in the work of Knoch~\cite{knoch1993structure} it was shown that the magnetic order is ferromagnetic at around 4.2~K for this compound. To understand this apparent contradiction with our results, we performed an additional low-temperature annealing (600~\degreecelsius) of our single crystal to study how such heat treatment can affect the magnetic state of Nd$_6$Fe$_{13}$Cu phase. The inserts in Fig.~\ref{MTMH5}(c,d) show the $M(H)$ dependencies measured on additionally annealed single crystals. Certainly, such annealing slightly changes the magnetic properties of the Nd$_6$Fe$_{13}$Cu single crystals, and a scanty ‘ferromagnetic’-like kink appears at the low field part of the $M(H)$ curves measured at 50~K and 10~K. This can be ascribed to the appearance of the local areas in the sample where antiferromagnetic order is not fully compensated and the large angle canted magnetic structure with the non-zero net magnetic moment is presented in the single crystal. All of this indicates that the Nd$_6$Fe$_{13}$Cu phase is not very stable at temperatures around $600\degreecelsius$ and this fact must be taken into account when considering the coercivity mechanism of Nd-Fe-B magnets containing 6:13:1 phase.

\begin{figure}[h!]
\centering
\includegraphics[width=0.8\textwidth]{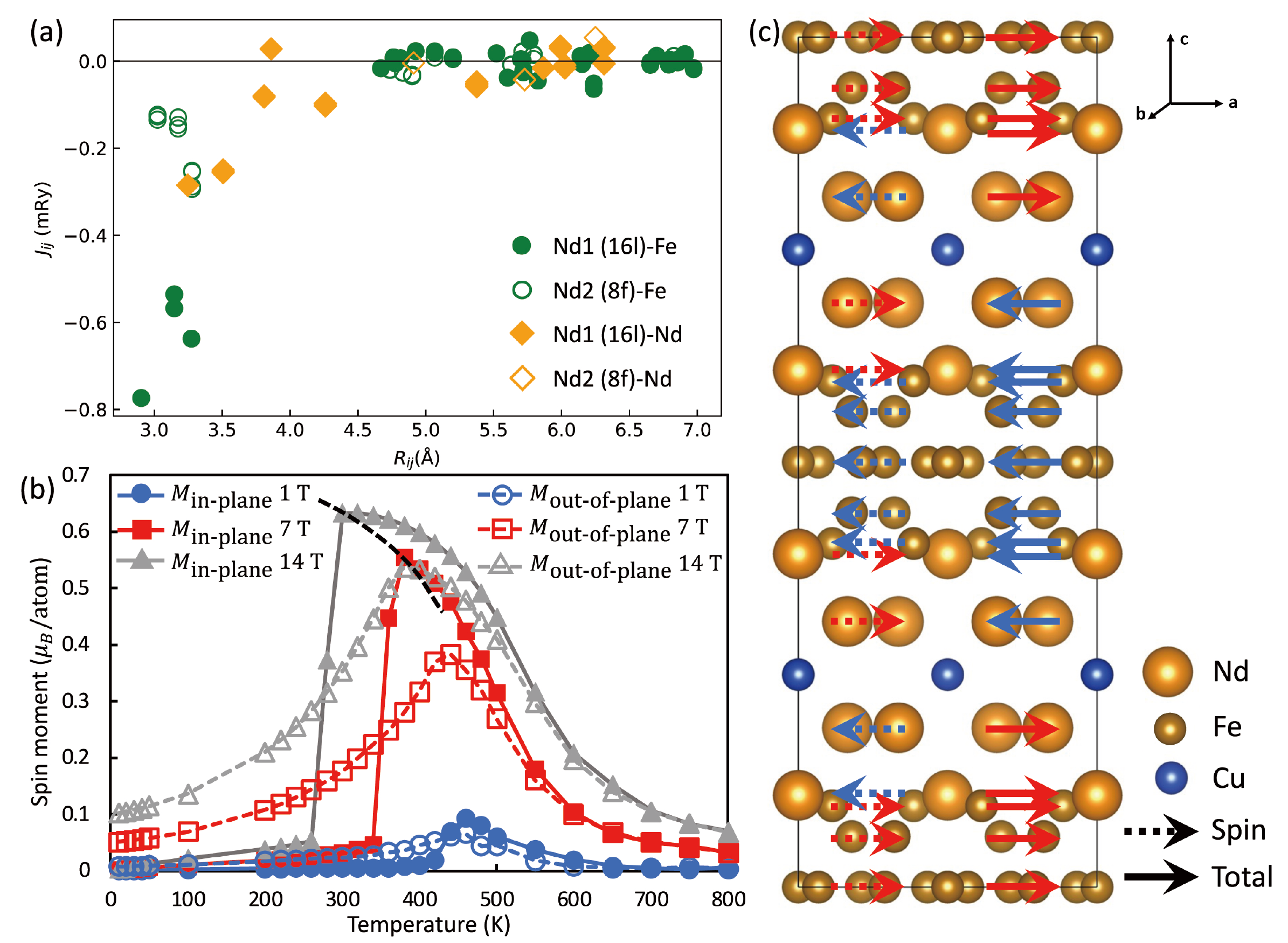}

\caption{(a) Exchange coupling parameters $J_{ij}$ as a function of atomic distance $R_{ij}$ in Nd$_6$Fe$_{13}$Cu. The $J_{ij}$s between the dominating atom pairs of Nd1(16l)-Fe, Nd2(8f)-Fe, Nd1(16l)-Nd and Nd2(8f)-Nd are demonstrated. 16l and 8f are the Wyckoff positions of Nd having coordinates of (0.164, 0.336, 0,313) and (0.0, 0.0, 0.108), respectively. (b) Evolution of averaged spin moments with respect to temperature under external out-of-plane/in-plane magnetic fields of 1 T, 7 T and 14 T, respectively. (c) Predicted ground state spin configuration (denoted by dashed arrows) and ground state magnetic configuration (denoted by solid arrows) if orbital moments are included.
}
\label{fig:mag_y}
\end{figure}

The ASD simulations are then performed to better understand the magnetic transition below $T_N$ under the external magnetic field along the a-axis (in-plane) magnetization direction. First, the $T_N$ is evaluated to be around 440~K based on the obtained magnetic susceptibility variation with temperature under no external magnetic field according to the ASD simulation results, which agrees rather well with the experiments ($T_N = 410$~K). Next, we employ ASD simulations under external magnetic fields of 1 T, 7 T and 14 T, respectively, to gain insights into the in-plane magnetic transitions below $T_N$ as illustrated in Fig.{~\ref{MTMH5}.}
 
In the ASD simulations, we assume easy-plane anisotropy as demonstrated by the experiments. The calculated exchange coupling parameters and the predicted ground-state spin configuration are displayed in Fig.~\ref{fig:mag_y} (a) and (c), respectively. As can be seen from Fig.~\ref{fig:mag_y} (c),  the Nd--Fe interface in the unit cell of Nd$_6$Fe$_{13}$Cu always prefers an AFM coupling (dashed arrows) between the spin moments, as suggested by the strongly negative $J_{ij}$ values for the first few nearest neighbours of Nd and Fe atoms (see Fig.~\ref{fig:mag_y} (a)). In addition, the two Nd layers below and above the Cu layers also prefer an AFM spin configuration, which can be attributed to the negative $J_{ij}$s between the Nd--Nd atom pairs. Note that the ASD simulations do not account for the orbital moments. According to the Hund's rule and the self-interaction-corrected relativistic DFT calculations~\cite{patrick2018rare}, the orbital moments of Nd atoms are antiparallel to ther spin moments, and the magnitude of the orbital moments is larger than that of the spin moments. Taking the orbital moments into consideration, the directions of the total magnetic moments corresponding to each atomic layer are illustrated using the solid arrows in Fig.~\ref{fig:mag_y} (c). 

\begin{figure}[h!]
\centering
\includegraphics[width=1\textwidth]{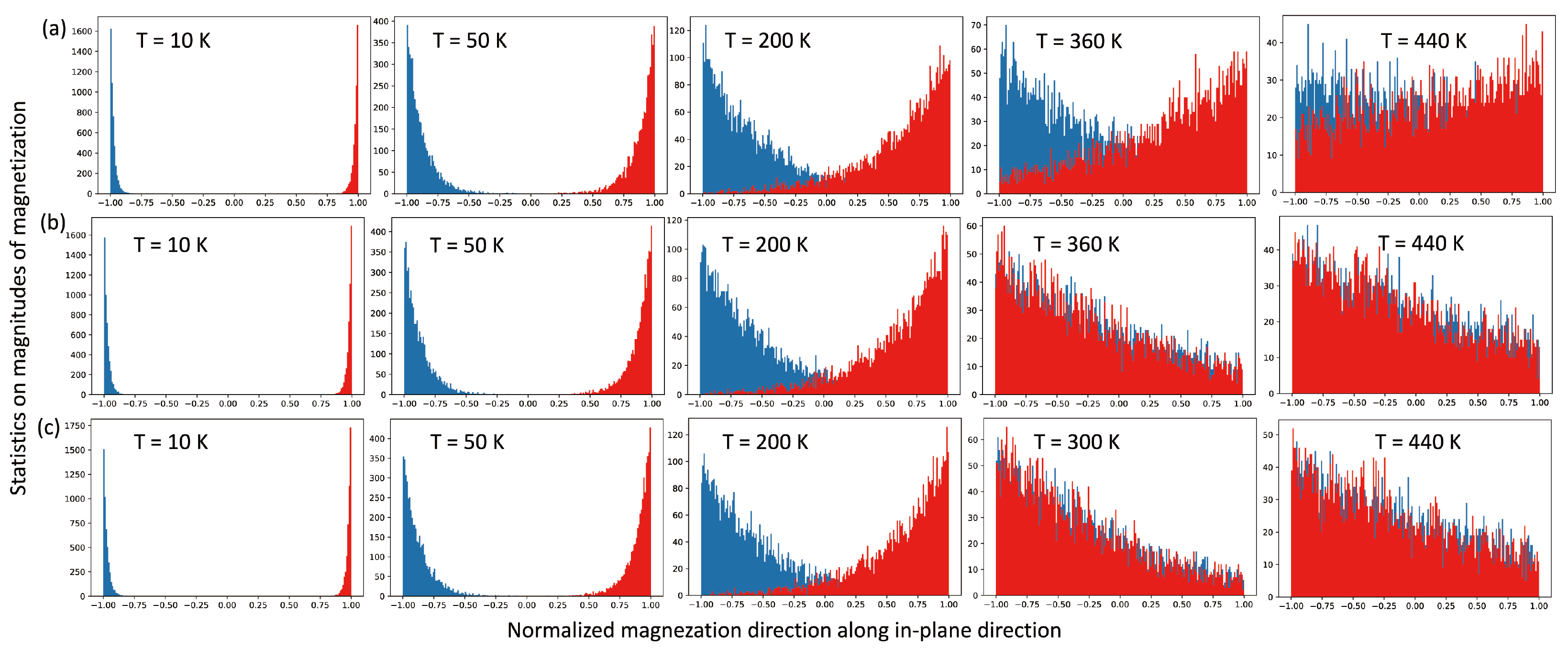}
\caption{Evolution of the distributions of Nd spin moments along the in-plane (a-axis) direction with temperature at external magnetic field equal to (a) 0 T, (b) 7 T and (c) 14 T.}
\label{fig:spin_hista}
\end{figure}

More importantly, according to the simulated spin moment curves under various temperatures with the magnetic fields aligned along the in-plane and out-of-plane directions (see Fig.~\ref{fig:mag_y} (b)), it is observed that there is a spin-flip transition as indicated by an abrupt increase of the spin moments when the external magnetic field is along the in-plane direction. A careful examination is then adopted in order to figure out the origin of such a spin flip. By plotting out the distributions of the in-plane spin moments at different temperatures and external magnetic fields, we find that with the facilitation of the external magnetic field, at $T=360$ K and $M_{in-plane}=7$ T, the Nd spin moments, which previously aligned along the direction of the external magnetic fields (atomic layers marked by red dashed arrows in Fig.~\ref{fig:mag_y} (c)), show an apparent tendency to point towards the opposite direction (see Fig.~\ref{fig:spin_hista} (b)). A similar transition happens at a lower temperature ($\sim$ 300 K) with a larger external magnetic field (14 T, see Fig.~\ref{fig:spin_hista} (c)). This phenomenon can be understood from the energy point of view. The increasing temperature and magnetic field provide an extra driving force to be competitive with the antiferromagnetic exchange couplings between the two Nd layers below and above the Cu layers. 

It is noted that it is challenging for the ASD simulations to reproduce the exact experimental measurements. For instance, with a magnetic field of 7 T at 10 K (see Fig.~\ref{MTMH5}), the spin-flip already occurs, while under the same magnetic field, the ASD simulation captures such a process at a temperature slightly higher than 340 K. Such a discrepancy can be due to the following reasons. First, our ASD simulations utilize a spin Hamiltonian and accordingly predict spin moment evolution as a function of temperature and external magnetic fields, meaning that the contribution from the orbital moments is not explicitly incorporated~\cite{kaplan1963theory}. Second, in this work, we assume an in-plane magnetic anisotropy constant favoring magnetization along the $y$ direction of -0.05 mRy for both Fe and Nd elements. We have to admit that this assumption is quantitatively rough. However, evaluating the element-resolved magnetic anisotropy based on \textit{ab initio} methods is challenging in rare-earth-transition-metal-based systems, not to mention its temperature dependence~\cite{patrick2019temperature}. In addition, the exchange coupling parameters are also calculated at 0 K, in which the temperature effect is missing as well.
Nevertheless, the qualitatively good description of the magnetic field and temperature effects can provide already insightful understanding on the magnetic properties of the 6:13:1 phase, consistent with the experimental observations.

To sum up, single crystals of Nd$_6$Fe$_{13}$Cu intermetallic compound were grown by the reactive flux technique. It is found that in fields below 6~T, Nd$_6$Fe$_{13}$Cu has an antiferromagnetic structure with a N\'eel temperature of 410~K. The magnetic fields can significantly affect the magnetic state of the single crystals, as below 150~K an external magnetic field of approx. 6~T applied along the a-axis induces a spin-flop transition, which indicates the competition between the exchange interactions of the Nd and Fe sublattices and their single-ion magnetocrystalline anisotropy \cite{utesov2018spiral,neel1952antiferromagnetism}. When the magnetic field is applied along the c-axis (hard axis of magnetization), no spin-flop transition is observed up to 14~T, which is typical for classical two-sublattice collinear antiferromagnets with relatively weak magnetic anisotropy \cite{bogdanov2007spin,yang2019spin}. 

In the 6:13:1 structure, the Fe, Nd and Cu layers are stacked in the direction of the c-axis and the directions of Fe and Nd moments change through the Cu layer~\cite{kajitani1992microstructure} \cite{iranmanesh2011magnetoelastic}. Atomistic spin dynamics simulations confirm that the Nd spin moments are aligned along the a-axis (in-plane), but an increase in temperature and/or magnetic field can significantly change the antiferromagnetic coupling between the two Nd layers below and above the Cu layers, which, in turn, is the reason for observed spin-flop transition.

Our results are valuable for the understanding of the coercivity mechanism in Cu-doped Nd-Fe-B magnets, where the Nd$_6$Fe$_{13}$Cu phase is a grain boundary phase and the microcrystalline grain of the matrix Nd$_2$Fe$_{14}$B phase must be exchange coupled~\cite{akdogan2014superferrimagnetism,murakami2014magnetism}. However, the 6-13-1 phase would hardly provide sufficient exchange-bias effect at the grain boundary~\cite{nayak2015design} or operate as a shell-antiferromagnet phase~\cite{scheibel2017room} due to the strong competition between its antiferromagnetic and ferromagnetic interactions. For this reason, it is very unlikely that the 6:13:1 phase enhances the coercivity of the Nd-Fe-Cu-B magnets.

The authors acknowledge the Deutsche Forschungsgemeinschaft (DFG, German Research Foundation), Project ID No. 405553726, TRR 270, subprojects A10, A05 and A01. The authors also thank the Lichtenberg high-performance center of TU Darmstadt for providing computational resources.

\bibliographystyle{elsarticle-num} 

\bibliography{cas-refs}

\providecommand{\noopsort}[1]{}\providecommand{\singleletter}[1]{#1}%
\begin{thebibliography}{10}
\expandafter\ifx\csname url\endcsname\relax
  \def\url#1{\texttt{#1}}\fi
\expandafter\ifx\csname urlprefix\endcsname\relax\def\urlprefix{URL }\fi
\expandafter\ifx\csname href\endcsname\relax
  \def\href#1#2{#2} \def\path#1{#1}\fi

\bibitem{gutfleisch2000controlling}
O.~Gutfleisch, Controlling the properties of high energy density permanent
  magnetic materials by different processing routes, Journal of Physics D:
  Applied Physics 33~(17) (2000) R157.

\bibitem{skokov2018heavy}
K.~Skokov, O.~Gutfleisch, Heavy rare earth free, free rare earth and rare earth
  free magnets-vision and reality, Scripta Materialia 154 (2018) 289--294.

\bibitem{sugimoto2011current}
S.~Sugimoto, Current status and recent topics of rare-earth permanent magnets,
  Journal of Physics D: Applied Physics 44~(6) (2011) 064001.

\bibitem{sawatzki2018anisotropic}
S.~Sawatzki, T.~Schneider, M.~Yi, E.~Bruder, S.~Ener, M.~Sch{\"o}nfeldt,
  K.~G{\"u}th, B.-X. Xu, O.~Gutfleisch, {Anisotropic local hardening in
  hot-deformed Nd-Fe-B permanent magnets}, Acta Materialia 147 (2018) 176--183.

\bibitem{schafer2022novel}
L.~Sch{\"a}fer, K.~Skokov, F.~Maccari, I.~Radulov, D.~Koch, A.~Mazilkin,
  E.~Adabifiroozjaei, L.~Molina-Luna, O.~Gutfleisch, {A Novel Magnetic
  Hardening Mechanism for Nd-Fe-B Permanent Magnets Based on Solid-State Phase
  Transformation}, Advanced Functional Materials (2022) 2208821.

\bibitem{liu2015grain}
J.~Liu, H.~Sepehri-Amin, T.~Ohkubo, K.~Hioki, A.~Hattori, T.~Schrefl, K.~Hono,
  {Grain size dependence of coercivity of hot-deformed Nd--Fe--B anisotropic
  magnets}, Acta Materialia 82 (2015) 336--343.

\bibitem{opelt2021upscaling}
K.~Opelt, T.~Ahmad, O.~Diehl, M.~Sch{\"o}nfeldt, E.~Brouwer, I.~Vogel, J.~D.
  Rossa, J.~Gassmann, S.~Ener, O.~Gutfleisch, {Upscaling the 2-Powder Method
  for the Manufacturing of Heavy Rare-Earth-Lean Sintered didymium-Based
  Magnets}, Advanced Engineering Materials 23~(10) (2021) 2100459.

\bibitem{sepehri2019development}
H.~Sepehri-Amin, I.~Dirba, X.~Tang, T.~Ohkubo, T.~Schrefl, O.~Gutfleisch,
  K.~Hono, {Development of high coercivity anisotropic Nd-Fe-B/Fe nanocomposite
  powder using hydrogenation disproportionation desorption recombination
  process}, Acta Materialia 175 (2019) 276--285.

\bibitem{schafer2021design}
L.~Sch{\"a}fer, K.~Skokov, J.~Liu, F.~Maccari, T.~Braun, S.~Riegg, I.~Radulov,
  J.~Gassmann, H.~Merschroth, J.~Harbig, et~al., {Design and Qualification of
  Pr--Fe--Cu--B Alloys for the Additive Manufacturing of Permanent Magnets},
  Advanced Functional Materials 31~(33) (2021) 2102148.

\bibitem{goll2021additive}
D.~Goll, F.~Trauter, T.~Bernthaler, J.~Schanz, H.~Riegel, G.~Schneider,
  {Additive manufacturing of bulk nanocrystalline Nd-Fe-B based permanent
  magnets}, Micromachines 12~(5) (2021) 538.

\bibitem{li2009role}
W.~Li, T.~Ohkubo, T.~Akiya, H.~Kato, K.~Hono, {The role of Cu addition in the
  coercivity enhancement of sintered Nd-Fe-B permanent magnets}, Journal of
  Materials Research 24~(2) (2009) 413--420.

\bibitem{herbst1991r}
J.~Herbst, {R2Fe14B materials: Intrinsic properties and technological aspects},
  Reviews of Modern Physics 63~(4) (1991) 819.

\bibitem{shimoda1988high}
T.~Shimoda, K.~Akioka, O.~Kobayashi, T.~Yamagami, {High-energy cast Pr-Fe-B
  magnets}, Journal of applied physics 64~(10) (1988) 5290--5292.

\bibitem{xu2019comparison}
X.~Xu, H.~Sepehri-Amin, T.~Sasaki, M.~Soder{\v{z}}nik, X.~Tang, T.~Ohkubo,
  K.~Hono, {Comparison of coercivity and squareness in hot-deformed and
  sintered magnets produced from a Nd-Fe-B-Cu-Ga alloy}, Scripta Materialia 160
  (2019) 9--14.

\bibitem{sasaki2016formation}
T.~Sasaki, T.~Ohkubo, Y.~Takada, T.~Sato, A.~Kato, Y.~Kaneko, K.~Hono,
  {Formation of non-ferromagnetic grain boundary phase in a Ga-doped Nd-rich
  Nd--Fe--B sintered magnet}, Scripta Materialia 113 (2016) 218--221.

\bibitem{huang2022enormous}
Q.~Huang, Q.~Jiang, Y.~Shi, S.~U. Rehman, X.~Wei, Z.~Li, D.~Shi, D.~Xu,
  Z.~Zhong, {Enormous improvement of the coercivity of Ga and Cu co-doping
  Nd-Fe-B sintered magnet by post-sinter annealing}, Journal of Alloys and
  Compounds 894 (2022) 162418.

\bibitem{zhang2022coercivity}
T.~Zhang, S.~Liu, Z.~Wu, L.~Zhang, R.~Yu, {Coercivity and remanence enhancement
  in hot-deformed Nd-Fe-B magnets by high-temperature short-term annealing
  process}, Journal of Alloys and Compounds 903 (2022) 163975.

\bibitem{zhang2022effect}
T.~Zhang, W.~Xing, F.~Chen, L.~Zhang, R.~Yu, {Effect of metallic Nd on the
  decomposition behavior of Nd6(Fe,Co)13Ga phase in hot-deformed Nd-Fe-B
  magnet}, Intermetallics 145 (2022) 107558.

\bibitem{hu1992crystal}
B.-p. Hu, J.~Coey, H.~Klesnar, P.~Rogl, {Crystal structure, magnetism and 57Fe
  M{\"o}ssbauer spectra of ternary RE6Fe11Al3 and RE6Fe13Ge compounds}, Journal
  of magnetism and magnetic materials 117~(1-2) (1992) 225--231.

\bibitem{yan1994magnetic}
Q.~Yan, P.~Zhang, X.~Sun, B.~Hu, Y.~Wang, X.~Rao, G.~Liu, C.~Gou, D.~Chen,
  Y.~Cheng, {The magnetic structure of Nd6Fe13Si}, Journal of Physics:
  Condensed Matter 6~(16) (1994) 3101.

\bibitem{kajitani1992microstructure}
T.~Kajitani, K.~Nagayama, T.~Umeda, {Microstructure of Cu-added Pr-Fe-B
  magnets: Crystallization of antiferromagnetic Pr6Fe13Cu in the boundary
  region}, Journal of magnetism and magnetic materials 117~(3) (1992) 379--386.

\bibitem{weitzer1994magnetism}
F.~Weitzer, A.~Leithe-Jasper, P.~Rogl, K.~Hiebl, A.~Rainbacher, G.~Wiesinger,
  W.~Steiner, J.~Friedl, F.~Wagner, {Magnetism of ternary compounds RE6Fe13X;
  RE= Pr, Nd; X= Cu, Ag, Au, Zn, Cd, and Hg}, Journal of applied physics
  75~(12) (1994) 7745--7751.

\bibitem{knoch1993structure}
K.~Knoch, A.~Le~Calvez, Q.~Qi, A.~Leithe-Jasper, J.~Coey, {Structure and
  magnetic properties of Nd6Fe13Cu}, Journal of applied physics 73~(10) (1993)
  5878--5880.

\bibitem{de1998magnetic}
C.~De~Groot, K.~Buschow, F.~De~Boer, {Magnetic properties of
  R$_6$Fe$_{13-x}$M$_{1+x}$ compounds and their hydrides}, Physical Review B
  57~(18) (1998) 11472.

\bibitem{hautot1998mossbauer}
D.~Hautot, G.~J. Long, F.~Grandjean, C.~De~Groot, K.~Buschow, {A M{\"o}ssbauer
  spectral study of Nd6Fe13X, where X is Cu, Ag, and Au and of the spin
  reorientation in Nd6Fe13Si}, Journal of applied physics 83~(3) (1998)
  1554--1562.

\bibitem{diop2018origin}
L.~Diop, M.~Kuz'Min, K.~Skokov, Y.~Skourski, O.~Gutfleisch, {Origin of
  field-induced discontinuous phase transitions in Nd2Fe17}, Physical Review B
  97~(5) (2018) 054406.

\bibitem{jonen1997metamagnetic}
S.~Jonen, H.~Rechenberg, {Metamagnetic transition in Nd$_6$Fe$_{14-x}$Al$_x$
  compounds}, Journal of applied physics 81~(8) (1997) 4054--4056.

\bibitem{zhao1995antiferromagnetic}
Z.~Zhao, C.~De~Groot, F.~De~Boer, K.~Buschow, {Antiferromagnetic interactions
  in Nd6Fe12Ga2-based compounds}, Physica B: Condensed Matter 211~(1-4) (1995)
  102--104.

\bibitem{de1996erratum}
C.~De~Groot, F.~De~Boer, K.~Buschow, D.~Hautot, G.~J. Long, F.~Grandjean,
  {Erratum: Magnetic and Mossbauer spectral properties of the compound
  Nd6Fe13Au}, Journal of alloys and compounds 245~(1-2) (1996) 188--188.

\bibitem{rosenberg1994metamagnetism}
M.~Rosenberg, R.~Zhou, M.~Velicescu, P.~Schrey, G.~Filoti, {Metamagnetism and
  spin arrangement in Nd6Fe13Sn}, Journal of Applied Physics 75~(10) (1994)
  6586--6588.

\bibitem{ruzitschka2002investigation}
R.~Ruzitschka, M.~Reissner, W.~Steiner, P.~Rogl, {Investigation of magnetic
  order in RE6Fe13X (RE=Nd, Pr; X= Pd, Sn, Si)}, Journal of magnetism and
  magnetic materials 242 (2002) 806--808.

\bibitem{iranmanesh2011magnetoelastic}
P.~Iranmanesh, N.~Tajabor, M.~R. Roknabadi, D.~Fruchart, F.~Pourarian,
  {Magnetoelastic properties of Nd6Fe13Cu intermetallic compound}, Physica B:
  Condensed Matter 406~(18) (2011) 3359--3362.

\bibitem{latturner2018clusters}
S.~E. Latturner, {Clusters, assemble: growth of intermetallic compounds from
  metal flux reactions}, Accounts of Chemical Research 51~(1) (2018) 40--48.

\bibitem{kanatzidis2005metal}
M.~G. Kanatzidis, R.~P{\"o}ttgen, W.~Jeitschko, The metal flux: a preparative
  tool for the exploration of intermetallic compounds, Angewandte Chemie
  International Edition 44~(43) (2005) 6996--7023.

\bibitem{eriksson2017atomistic}
O.~Eriksson, A.~Bergman, L.~Bergqvist, J.~Hellsvik, Atomistic spin dynamics:
  foundations and applications, Oxford university press, 2017.

\bibitem{terasawa2019efficient}
A.~Terasawa, M.~Matsumoto, T.~Ozaki, Y.~Gohda, Efficient algorithm based on
  liechtenstein method for computing exchange coupling constants using
  localized basis set, Journal of the Physical Society of Japan 88~(11) (2019)
  114706.

\bibitem{ozaki2003variationally}
T.~Ozaki, Variationally optimized atomic orbitals for large-scale electronic
  structures, Physical Review B 67~(15) (2003) 155108.

\bibitem{han2007magnetic}
M.~J. Han, T.~Ozaki, J.~Yu, {Magnetic ordering and exchange interactions in
  multiferroic GaFeO3}, Physical Review B 75~(6) (2007) 060404.

\bibitem{isnard2002neutron}
O.~Isnard, G.~J. Long, D.~Hautot, K.~Buschow, F.~Grandjean, {A neutron
  diffraction and M{\"o}ssbauer spectral study of the magnetic spin
  reorientation in Nd6Fe13Si}, Journal of Physics: Condensed Matter 14~(47)
  (2002) 12391.

\bibitem{patrick2018rare}
C.~E. Patrick, J.~B. Staunton, Rare-earth/transition-metal magnets at finite
  temperature: Self-interaction-corrected relativistic density functional
  theory in the disordered local moment picture, Physical Review B 97~(22)
  (2018) 224415.

\bibitem{kaplan1963theory}
T.~Kaplan, D.~Lyons, Theory of indirect exchange interactions in rare-earth
  metals, Physical Review 129~(5) (1963) 2072.

\bibitem{patrick2019temperature}
C.~E. Patrick, J.~B. Staunton, {Temperature-dependent magnetocrystalline
  anisotropy of rare earth/transition metal permanent magnets from first
  principles: the light RCo5 (R= Y, La-Gd) intermetallics}, Physical Review
  Materials 3~(10) (2019) 101401.

\bibitem{utesov2018spiral}
O.~Utesov, A.~Syromyatnikov, Spiral plane flops in frustrated helimagnets in
  external magnetic field, Physical Review B 98~(18) (2018) 184406.

\bibitem{neel1952antiferromagnetism}
L.~N{\'e}el, Antiferromagnetism and ferrimagnetism, Proceedings of the Physical
  Society. Section A 65~(11) (1952) 869.

\bibitem{bogdanov2007spin}
A.~Bogdanov, A.~Zhuravlev, U.~R{\"o}{\ss}ler, Spin-flop transition in uniaxial
  antiferromagnets: Magnetic phases, reorientation effects, and multidomain
  states, Physical Review B 75~(9) (2007) 094425.

\bibitem{yang2019spin}
J.~Yang, W.~Yang, Q.~Zhu, B.~Chen, H.~Wang, Q.~Mao, J.~Du, Z.~Lou, M.~Fang,
  {Spin-flop transition and magnetic phase diagram in CsCo2Se2 revealed by
  torque and resistivity measurements}, Journal of Magnetism and Magnetic
  Materials 474 (2019) 70--75.

\bibitem{akdogan2014superferrimagnetism}
O.~Akdogan, A.~Dobrynin, D.~Le~Roy, N.~Dempsey, D.~Givord, {Superferrimagnetism
  in hard Nd-Fe-B thick films, an original concept for coercivity enhancement},
  Journal of Applied Physics 115~(17) (2014) 17A764.

\bibitem{murakami2014magnetism}
Y.~Murakami, T.~Tanigaki, T.~Sasaki, Y.~Takeno, H.~Park, T.~Matsuda, T.~Ohkubo,
  K.~Hono, D.~Shindo, {Magnetism of ultrathin intergranular boundary regions in
  Nd--Fe--B permanent magnets}, Acta Materialia 71 (2014) 370--379.

\bibitem{nayak2015design}
A.~K. Nayak, M.~Nicklas, S.~Chadov, P.~Khuntia, C.~Shekhar, A.~Kalache,
  M.~Baenitz, Y.~Skourski, V.~K. Guduru, A.~Puri, et~al., {Design of
  compensated ferrimagnetic Heusler alloys for giant tunable exchange bias},
  Nature materials 14~(7) (2015) 679--684.

\bibitem{scheibel2017room}
F.~Scheibel, D.~Spoddig, R.~Meckenstock, T.~Gottschall, A.~{\c{C}}ak{\i}r,
  T.~Krenke, M.~Farle, O.~Gutfleisch, M.~Acet, Room-temperature five-tesla
  coercivity of a rare-earth-free shell-ferromagnet, Applied Physics Letters
  110~(19) (2017) 192406.

\end{thebibliography}

\end{document}